\documentclass[twocolumn,a4paper,10pt]{article}
\usepackage{graphicx}
\usepackage{txfonts}
\begin{document}
\twocolumn[\Large\begin{center}
{\bf Recent Results from Telescope Array}\\
\vspace*{3mm} \large
M. Fukushima for the Telescope Array Collaboration\\
Institute for Cosmic Ray Research, The University of Tokyo,
Kashiwanoha 5-1-5,\\
Kashiwa, Chiba, 277-8582 Japan\\
\vspace*{5mm}
\end{center} ]

%%%%%%%% -----------------------------------------------------
\section*{Abstract}
The Telescope Array (TA) is an experiment to observe 
Ultra-High Energy Cosmic Rays (UHECRs).
TA's recent results, the energy spectrum and anisotropy
based on the 6-year surface array data, and 
the primary composition obtained from
the shower maximum (X$_{\rm MAX}$) are reported.
The spectrum demonstrates a clear dip and cutoff.
The shape of the spectrum is well described by the energy
loss of extra-galactic protons interacting with the cosmic
microwave background (CMB). Above the cutoff, 
a medium-scale (20$^\circ$ radius) flux enhancement was
observed near the Ursa-Major.  
A chance probability of creating this hotspot
from the isotropic flux is 4.0 $\sigma$. 
The measured $<$X$_{\rm MAX}$$>$ is consistent
with the primary being proton or light nuclei
for energies 10$^{18.2}$~eV - 10$^{19.2}$~eV.
This report is to appear in the proceedings of
the ISVHECRI-2014 symposium at CERN, 
August 18th-22nd, 2014. 

%%%%%%%% -----------------------------------------------------
\section{Introduction}
%\label{introduction}
A cutoff of the cosmic ray energy spectrum around 10$^{20}$~eV was
suggested by Greisen, Zatsepin and Kuzmin (GZK) in 1966
\cite{gzk} just
after the discovery of the cosmic microwave background (CMB).  
When the energy of cosmic ray proton approaches to ultra-high
energies, pion photo-production with the CMB starts. The creation
and decay of the nucleon resonance deprives a significant
energy from the propagating proton. This process
produces a cutoff and a horizon
for UHECRs arriving at the Earth.
Precise measurements of the pion photo-production cross section
by  accelerator experiments, and of the CMB radiation
temperature by  space experiments allowed a precise
prediction for the cutoff energy and the horizon,
assuming certain models for  cosmic ray production and
its cosmological development
\cite{gzk-energy}. 
Thus, a search for the GZK cutoff became an important subject
for  cosmic ray experiments, and a search for 
UHECR sources within the GZK horizon became
a realistic hope.

Searches for the GZK effect by ground based
air shower arrays have been
associated with the experimental difficulties
such as the very low arrival rate,
one event per year on an area of 100 km$^2$ at $\sim$10$^{20}$~eV, 
and large ambiguities in the determination of primary
energy and composition by the sparse sampling of
shower particles arriving on the ground.

A new air fluorescence technique for measuring  UHECRs
was established by the Fly's Eye (FE)
and High Resolution FE (HiRes) 
\cite{fy-experiment,hires-experiment} by 2000.
In this technique, the longitudinal development of air shower
in the atmosphere was recorded by fast imaging telescopes,
and the primary cosmic ray energy was determined
by integrating the profile of the
shower development. A stochastic measurement 
of the primary composition
became possible by observing the average and fluctuation
of X$_{\rm MAX}$ over many shower samples. 

The Telescope Array (TA) employs 
an array of Surface Detectors (SDs)
covering a large ground area of 700~km$^2$,
together with a set of Fluorescence Detectors
(FDs) overlooking the SD area
\cite{ta-experiment-sd,ta-experiment-fd}. It is 
installed in the West Desert of Utah, USA, 
and has been taking data since 2008. 
Whenever possible the SDs and FDs make
simultaneous measurement of the event, however
the operation of FD is limited to  moonless clear nights
and its duty factor is at a level of 10\%. 
The SD runs with close to 100\% duty, collects
higher statistics, and its sky sampling
is uniform in equatorial longitude. 
A similar type of 
hybrid experiment was installed
at the Pierre Auger Observatory (PAO)
in the Argentinian pampas
\cite{auger-experiment}, 
and has been taking data since 2004
with 3000~km$^2$ SD array.

The search for the GZK effect, and understanding 
the nature of UHECRs is rapidly proceeding with
the two large-scale hybrid experiments, TA and PAO. 
We report here the recent results from TA
as of  summer 2014.

%%%%%%%% -----------------------------------------------------
\section{TA Experiment}
\begin{figure}[tbh]
\begin{center}
  \includegraphics[width=0.85\linewidth]
  {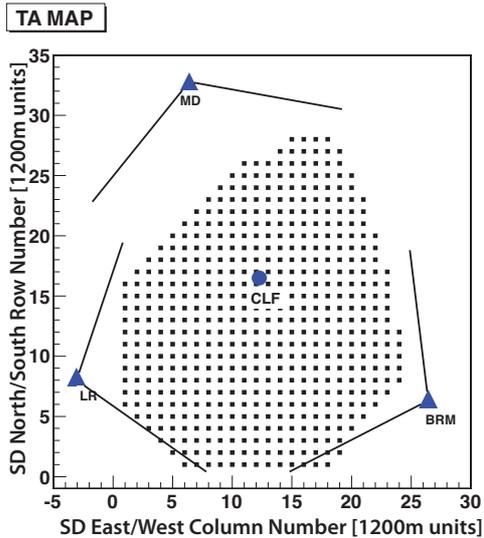}
  \caption{Detector layout of the TA experiment.
Filled black squares indicate
the locations of the SDs, and triangles are 
the FD stations. The Central Laser Facility (CLF) is 
shown by the circle at the center of the array.}
  \label{ta-map.label}
\end{center}
\end{figure}

The TA is located in Utah, USA at a latitude of 39.3$^\circ$ North and
a longitude of 112.9$^\circ$ West. The altitude
is approximately 1400 m above the sea level. The TA encloses 
507 SDs, 3 stations of FDs and several calibration devices on site
(see Fig.\ref{ta-map.label}).

The TA SDs are deployed in a 1.2 km grid covering the ground area of
700~km$^2$. Each SD is made of 2 layers of 1.2 cm thick, 3 m$^2$ large
plastic scintillaters overlaid on top of each other
\cite{ta-experiment-sd}. 
Each layer is read out independently with a PMT
using wavelength shifter fibers.
The waveform is locally recorded 
by 12bit, 50MHz FADC for any hit with more than
0.3 Minimum Ionizing Particles (MIPs)
in both  scintillators in coincidence.
These events are also locally histogrammed and used
for the calibration. The stored waveforms are read out
using a 2.4~GHz wireless network when more than 3 MIPs are recorded
in 3 adjacent SDs within 8 $\mu$s.

The TA FD is composed of 38 fluorescence telescopes installed in
3 FD stations at Black Rock Mesa (BRM), Long Ridge (LR) and Middle Drum
(MD) surrounding the SD array. 
The FD at MD station, FD(MD), is instrumented with the
refurbished HiRes telescopes with 5.1m$^2$ spherical mirror, and
the FD(BRM) and FD(LR) are newly designed telescopes
with 6.8~m$^2$ spherical mirror
\cite{ta-experiment-fd}.
Each FD station monitors the night sky from
3$^\circ$ to 31$^\circ$ (MD) or 33$^\circ$ (BRM, LR)
in the elevation, and 112$^\circ$(MD) or
108$^\circ$(BRM, LR) in the azimuth.
The readout pixel of PMT camera
is approximately 1.0$^\circ$ cone
for all 3 stations. The PMT signal is recorded by the
HiRes-1 sample and hold electronics at the MD station, and
by a 10 MHz, 14-bit equivalent FADCs at the BRM and LR stations.
A Central Laser Facility (CLF) is installed at the equidistant
point from the 3 FD stations and regularly shoots a YAG laser 
into the sky for atmospheric monitoring. 
An Electron Light Source (ELS), a 40 MeV electron linac,
is installed 100~m in front of the FD(BRM)
to make an end-to-end calibration
by shooting an electron beam into the sky.

%%%%%%%% -----------------------------------------------------
\section{Energy Spectrum}
\begin{figure}[tbh]
\begin{center}
  \includegraphics[width=0.85\linewidth]
  {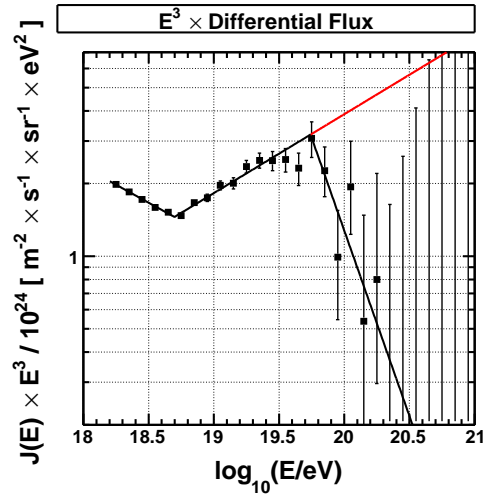}
  \caption{Energy spectrum obtained from 
the 6-year TA SD data. Solid black lines show
the broken power law fit to the data. The red line
is the extrapolation of the central segment to higher energies. 
}
  \label{spectrum-sd-6y.label}
\end{center}
\end{figure}

The E$^3$ multiplied differential flux
measured by the TA SD is plotted in Fig.\ref{spectrum-sd-6y.label}
where E is the energy of the primary cosmic rays. 
The spectrum is composed of 17,763 events above 10$^{18.2}$~eV
collected in 6 years of SD operation from May 2008 to May 2014.
The corresponding exposure is 5400 km$^2$ sr year
above 10$^{18.8}$~eV.
\begin{figure}[tbh]
\begin{center}
  \includegraphics[width=0.85\linewidth]
  {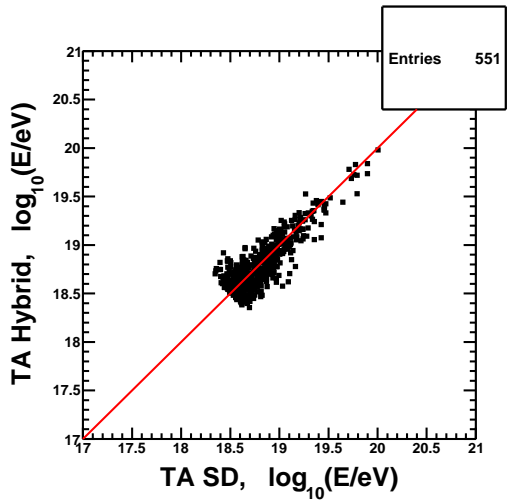}
  \caption{Scatter plot of E$_{\rm SD}$/1.27 (abscissa)
and E$_{\rm FD}$ (ordinate) for 551 hybrid events
taken for May 2008 - May 2013. The red line corresponds
to the equality of the rescaled E$_{\rm SD}$ and E$_{\rm FD}$.
}
  \label{spectrum-efd-esd-scatterplot.label}
\end{center}
\end{figure}

The analysis of the SD data
is described in our previous publication \cite{ta-cutoff}.
For each SD event, 
the shower core location
and the arrival direction were determined by
fitting the shower lateral distribution and the SD hit timings
with empirical formulae. For the spectrum measurement, the 
following cuts were applied to select events:
5 or more SDs, zenith angle $\theta$
less than 45$^\circ$, the core location more than
1200~m inside the array boarder,
and the fitting $\chiup^2$ and uncertainty estimates
of energy and angle are
within certain limits \cite{ta-cutoff}.
The energy of the reconstructed event
was determined by comparing the observed particle density
800~m away from the shower axis, S(800), with 
the expected particle density estimated by 
the air shower simulation. 
The CORSIKA and GEANT4 simulation tools and the
de-thinning method were used
for the primary proton to produce
large numbers of air shower particles
on the ground \cite{ta-simulation}.
The hadronic interaction model QGSJET-II-03
was used in the air shower generation. 

The obtained SD energy, E$_{\rm SD}$, is then
compared with E$_{\rm FD}$, the energy measured by the FD
for SD-FD simultaneously measured (hybrid)
events. The scatter plot of E$_{\rm SD}$/1.27
and E$_{\rm FD}$ is shown in
Fig.\ref{spectrum-efd-esd-scatterplot.label}
for energies above 10$^{18.5}$~eV.
As seen in the Figure, a simple rescaling of  
E$_{\rm SD}$ restores a good equality
between two measured energies.

The aperture and the
resolution of the accepted events were
calculated by generating MC events
according to the measured
spectrum, and applying exactly the same analysis
and the selection criteria as the data.
Distributions of simulated events for S(800),
core location, energy, zenith and azimuth angles were
checked for consistency with the distributions
of data events. The acceptance
is about 10\% at 10$^{18.2}$~eV
and rises to 100\% above 10$^{18.8}$~eV. 
The energy resolution
above 10$^{19}$~eV is 20\% and
the directional resolution 
is 1.0$^\circ$ above 10$^{19.7}$~eV
\cite{ta-cutoff, ta-hotspot}.

As seen in Fig.\ref{spectrum-sd-6y.label}, 
the energy spectrum is well fitted 
by a broken power law (BPL) demonstrating
the structure of ``ankle'' and ``cutoff''.
These structures were first observed by the HiRes
experiment\cite{hires-cutoff}
in 2008. The PAO also saw a strong flux suppression
around this energy in 2008 \cite{auger-cutoff}.
The fitted BPL parameters
are listed in
Table\ref{spectrum-fit-parameters.table}.
The energy E$_{1/2}$, where
the integral flux with cutoff becomes lower
by a factor of 2 than the power-law extrapolation,
is 10$^{19.73}$ eV for the data, and it agrees
well with the prediction of 
the GZK cutoff (10$^{19.72}$ eV)
by Berezinsky, Gazizov and Grigorieva  
\cite{gzk-energy}.
\begin{table}[htb]
\begin{center}
  \caption{Results of the BPL fit}
  \vspace{3mm}
  \begin{tabular}{clr}
    E$_1$      & 1st break energy (ankle)            & 10$^{18.70 \pm 0.02}$ eV \\
    E$_2$      & 2nd break energy (cutoff)           & 10$^{19.74 \pm 0.04}$ eV \\
    E$_{1/2}$  & GZK energy (\cite{gzk-energy})      & 10$^{19.73 \pm 0.04}$ eV \\
    $\gamma_1$ & power index for E $<$ E$_1$           & -3.30 $\pm$ 0.03 \\
    $\gamma_2$ & power index for E$_1$ $<$ E $<$ E$_2$ & -2.67 $\pm$ 0.03 \\
    $\gamma_3$ & power index for E$_2$ $<$ E           & -4.54 $\pm$ 0.44 \\
   \end{tabular}
  \label{spectrum-fit-parameters.table}
\end{center}
\end{table}

Above the energy E$_2$, 32 events are observed, whereas 85.9 events are
expected if the second break (cutoff) at E$_2$ does not exist.
It signifies a~6.6 $\sigma$ deviation from the continued
spectrum without the cutoff.
\begin{figure}[tbh]
\begin{center}
  \includegraphics[width=1.2\linewidth]
  {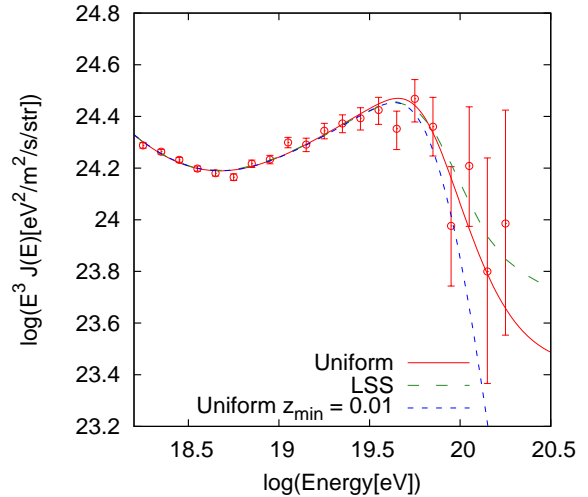}
  \caption{TA's 5-year energy spectrum fitted with the
simulation of extra-galactic protons.
Three cases of calculation are shown: (Uniform) uniform 
cosmic ray source distribution as described in the text,
(LSS) source distribution according to the
2MASS Galaxy Redshift Catalog (XSCz) up to 250~Mpc,
and uniform beyond 250~Mpc, 
(Uniform z$_{\rm min}$=0.01) uniform source distribution
for 0.01~$<$~z, but no source for z~$<$~0.01. 
In plotting this Figure, the best fit energy
shift $\Delta E$ is not applied to the data,
instead $-\Delta E$ is applied to the simulated spectrum. 
}
  \label{kido-fit.label}
\end{center}
\end{figure}
\begin{figure}[tbh]
\begin{center}
  \includegraphics[width=1.2\linewidth]
  {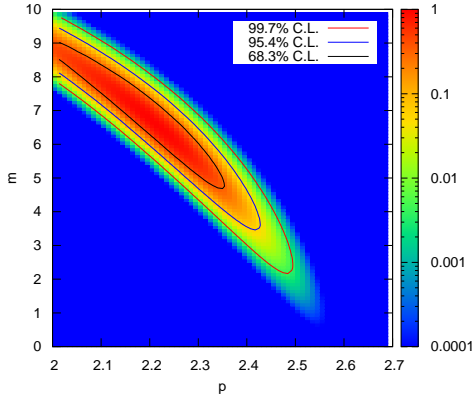}
  \caption{Permitted region of the source power index, p,
and the evolution parameter, m, of uniformly
distributed cosmic ray proton sources in z~$<$~0.7.}
  \label{kido-pm.label}
\end{center}
\end{figure}

The observed spectrum is fitted with the expectation
from the cosmic ray transport calculation and simulation
\cite{ta-spectrum-fit}.
The rate of cosmic ray production
at redshift $z$ is modeled as 
\begin{equation}
Q(E,z) = \alpha~E^{-p}~(1+z)^{(3+m)},~~E<E_{max},~z<z_{max}
\nonumber
\end{equation}
where p and m are the spectral index and the
cosmological evolution parameter.
The values of E$_{\rm max} = 10^{21}$~eV 
and $z_{\rm max}$ = 2.0 were used
such that these parameters do
not affect the spectral shape in the fitting region.
The generated protons are transported rectilinearly
in the extra-galactic space, and energy losses caused
by the interactions with the cosmic microwave background and infra-red photons were calculated. The calculation
by the transport equation (TransportCR) and
the MC simulation (CRPropa) were compared and they agreed well
after some updates are applied
to the CRPropa parameters 
\cite{kalashev-kido}.

The observed spectrum was fitted by optimizing 4 parameters in
the transport calculation:
p, m, overall normalization ($\alpha$) and the
energy scale ($\Delta E$). For the simulated spectrum,
events were generated, transported with energy loss
and the energy smearing same as the data resolution
was applied before histogramming in the spectrum. 
For the data spectrum, only the statistical error
was considered in the fitting.
The best fit to the data spectrum was obtained
with p~=~2.2, m~=~6.4 and log$\Delta E~=~-0.05$ with
$\chiup^2/{\rm NDF}$ = $21.3/17$. It is shown in
Fig.\ref{kido-fit.label} as ``Uniform''
\footnote{
The value of log$\Delta E~=~-0.05$ means the best fit was
obtained when the measured energy is decreased by
10\%. Adding the flux systematic error of $\sim$3\% in quadrature
\cite{ivanov-phd-thesis}
changed the best fit to p~=~2.2,
m~=~6.7 and log$\Delta E~=~-0.03$
with $\chiup^2/{\rm NDF}$ = $12.4/17$.}.
The best fit energy scale log$\Delta E~=~-0.05$
is well within the systematic uncertainty of 21\%
for the TA hybrid events\cite{ikeda-hybrid}.

Fig.\ref{kido-fit.label} demonstrates that 
the TA 5-year SD spectrum
can be fitted well
with uniformly distributed, cosmologically evolving
cosmic ray proton sources without requiring
contributions from other components. The allowed region 
of the p and m are shown in 
Fig.\ref{kido-pm.label}. It should be noted
that most particles above 10$^{18.2}$~eV are coming from
z~$<$~0.7, and constraints in Fig.\ref{kido-pm.label}
apply only for such sources.

%%%%%%%% -----------------------------------------------------
\section{Energy Spectrum by TALE}
\begin{figure}[tbh]
\begin{center}
  \includegraphics[width=0.85\linewidth]
  {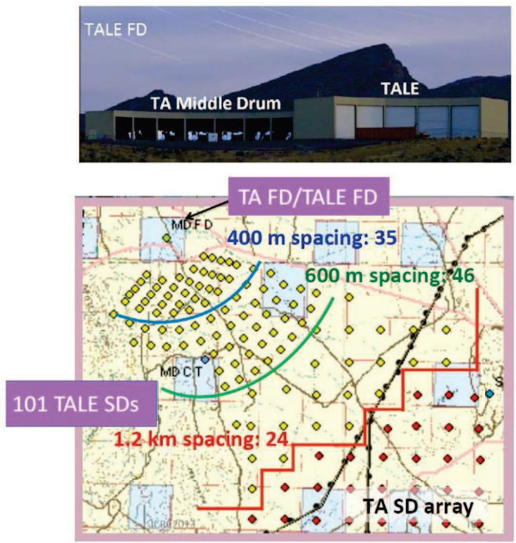}
  \caption{Detector layout of TALE. The TALE FD is
installed next to the FD(MD). The area between FD(MD)
and the TA SD will be filled with in-fill arrays.}
  \label{tale-map.label}
\end{center}
\end{figure}

The TA Low Energy extension, TALE, is a hybrid
addition to the TA
installed next to the north FD station of TA, FD(MD)
(see Fig.\ref{tale-map.label}).
It is composed of 10 additional FD telescopes refurbished from HiRes
observing the higher elevation of
31$^\circ$ $-$ 59$^\circ$ above the FoV of FD(MD).
The azimuthal coverage is $\sim$100$^\circ$.
An infill array of 105 SDs is located in between
the TALE FDs and the TA SD with 400~m, 600~m and 1200~m spacings.
The TALE SD works as an independent SD array as well
as the hybrid array with TALE FDs. 
The TALE FDs were commissioned in September 2013.
A total of 35 SDs were deployed and are being tested.
The TALE and TA together will be
measuring cosmic rays in a wide energy range of
10$^{16.5}$eV - 10$^{20.5}$ eV in a single experiment. 
The use of Cherenkov analysis in TALE FD may
decrease the energy threshold as low as 10$^{15.5}$~eV.
\begin{figure}[tbh]
\begin{center}
  \includegraphics[width=0.95\linewidth]
  {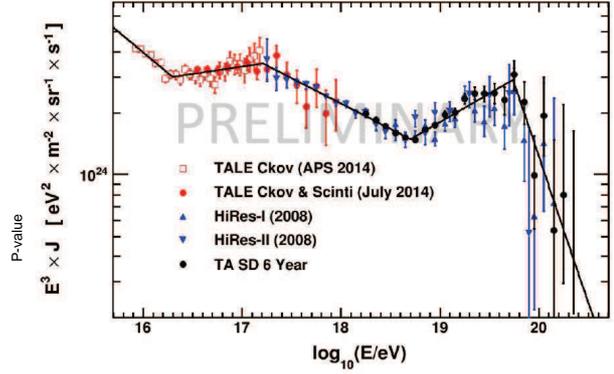}
  \caption{TA and TALE energy spectrum with
the BPL fit. TALE results in red open squares and
in closed red circles are preliminary.
HiRes data are from
\cite{hires-cutoff}.
}
  \label{tale-spectrum.label}
\end{center}
\end{figure}

In Fig.\ref{tale-spectrum.label}, the preliminary TALE
energy spectrum is plotted together with TA SD, 
HiRes-1 and HiRes-2 measurements.
The spectrum from 10$^{15.9}$~eV to 10$^{17.2}$~eV,
designated as TALE Ckov in
Fig.\ref{tale-spectrum.label}, is obtained
by using the TALE FD as an Imaging Atmospheric Cherenkov
Telescope (IACT)
\cite{tale-cherenkov-spectrum}.
The Cherenkov events seen by TALE FD
are with short duration (100 - 600~ns) and
short track often confined in one FD camera. They are
distinct from the fluorescence events, and are analyzed
as a FD monocular event. Its geometry is reconstructed
by the profile constrained geometry fit (PCGF) developed
for HiRes-1. The energy is obtained by using
the Corsika/IACT simulation
\cite{corsika-iact}.
The spectrum from 10$^{16.5}$~eV to 10$^{18}$~eV
are obtained by using the TALE FD as an air
fluorescence telescope.

The whole spectrum in Fig.\ref{tale-spectrum.label}
can be fitted by the BPL with two additional
breaks at 10$^{16.3}$eV and 10$^{17.2}$ eV.
Similar low energy features have been reported recently 
by KASCADE-Grande, Tibet AS$\gamma$, IceTop,
Tunka and other experiments
\cite{low-energy-spectrum}.

%%%%%%%% -----------------------------------------------------
\section{Composition}
Information on the primary cosmic ray composition
is obtained from the measurement
of X$_{\rm MAX}$
in the atmosphere, which is expected to scale
with $lnA$ and $lnE$ in average, where A is
the mass number of the cosmic ray nuclide:
\begin{equation}
<X_{\rm MAX}>~= C_\alpha~(lnE~- <lnA>)~+~C_\beta
\nonumber
\end{equation}

\begin{figure}[tbh]
\begin{center}
  \includegraphics[width=0.75\linewidth]
  {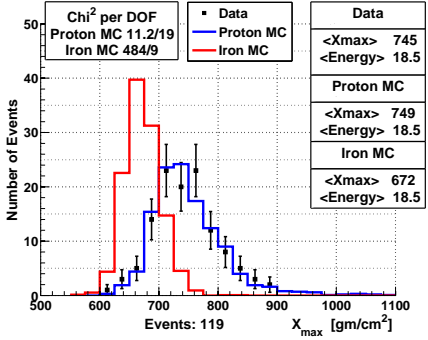}
  \caption{X$_{\rm MAX}$ distribution for energies
10$^{18.4}$eV $<$ E $<$ 10$^{18.6}$eV. 
Histograms are the results of MC simulations with QGSJET-II-03
hadronic interaction model: (Blue) iron and (Red) proton.
MCs are normalized to the same total number of events with
the data. Events are reconstructed as
the hybrid of FD(MD) and SD. The data much more
closely resembles  proton simulation
than  iron simulation.
}
  \label{xmax-distribution-184-186.label}
\end{center}
\end{figure}
\begin{figure}[tbh]
\begin{center}
  \includegraphics[width=0.75\linewidth]
  {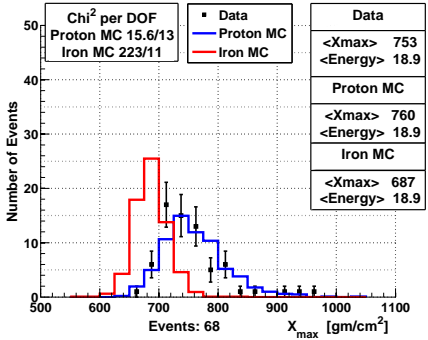}
  \caption{Same as Fig.\ref{xmax-distribution-184-186.label}
but for energies
10$^{18.8}$ eV $<$ E $<$ 10$^{19.0}$ eV.
The data looks very protonic as in
Fig.\ref{xmax-distribution-184-186.label}.
}
  \label{xmax-distribution-188-190.label}
\end{center}
\end{figure}

A most reliable way of measuring X$_{\rm MAX}$ is
the imaging of shower profile by the FD.
Results so far obtained on  X$_{\rm MAX}$
are by the stereo FD measurement from HiRes
\cite{hires-xmax},
the hybrid measurement from Auger
\cite{auger-xmax}, and
the stereo measurement from TA
\cite{ta-xmax-stereo}.
Each method corresponds to overlapping but specific
phase space; the hybrid analysis requires
the SD hits thus preferring more vertical events than the stereo,
the stereo analysis requires higher energy threshold
than the hybrid and so forth.
Our goal is to establish a consistent set of results 
in the X$_{\rm MAX}$ measurement among different analysis methods,
and to make sure the differences between experiments
are not due to the choice of different analysis methods.
We report here the results of the hybrid analysis
using the FD(MD) 
\cite{ta-xmax-md-hybrid}.

\begin{figure}[tbh]
\begin{center}
  \includegraphics[width=0.9\linewidth]
  {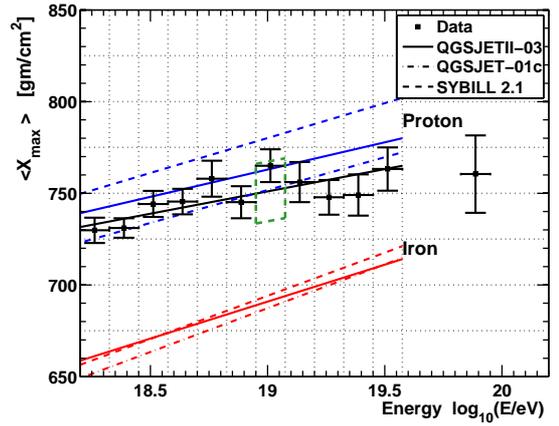}
  \caption{Elongation of average X$_{\rm MAX}$. Error bars are
statistical only. MC expectation using several interaction models
are shown for protons (blue) and irons (red). Both the data
and the simulation include reconstruction and selection biases. 
}
  \label{xmax-elongation.label}
\end{center}
\end{figure}

For this analysis, we selected events which were
triggered and reconstructed independently
by the monocular FD(MD) and by the SD.
The FD and SD reconstructions are then combined
into a hybrid analysis; the core location and impact
timing information from the SD is combined
with the timing information from the FD, thus
improving the geometry reconstruction of the
monocular FD. The shower profile
was then fitted with the Gaisser-Hillas function 
obtained from the pre-made CORSIKA shower library
to determine the best fit X$_{\rm MAX}$.
Pattern recognition cuts were developed
based on the experience of event eye scanning,
and were used to remove events
with poor fitting quality.

We started with the good weather cut passed, 
fully reconstructed 1916 events
for 10$^{18.2}$eV $<$ E collected in 5 years
of TA runs between May 2008 and May 2013.
After applying the
geometrical cuts: bracketed X$_{\rm MAX}$,
$\theta$ $<$ 56$^\circ$, 
and the SD core location inside the array,
843 events remained. The final sample passing
the pattern recognition cuts was 438 events.
The X$_{\rm MAX}$ resolution with pattern recognition cuts
were studied by simulation.
At 10$^{18.5}$ eV,
the X$_{\rm MAX}$ resolution 
was 35 g/cm$^2$ without pattern recognition cuts,
then improved to 23 g/cm$^2$ with cuts.
At 10$^{19.5}$ eV, the resolution was 15 g/cm$^2$
\cite{ta-xmax-md-hybrid}. 

The distributions of X$_{\rm MAX}$
for two energy regions
E $\sim$ 10$^{18.5}$eV and 
E $\sim$ 10$^{18.9}$eV
are shown in
Fig.\ref{xmax-distribution-184-186.label} and
Fig.\ref{xmax-distribution-188-190.label}
together with the MC simulation results
for protons and irons
\footnote{See \cite{ta-xmax-md-hybrid} 
for other energy regions.}.
The MC events went through the same reconstruction and
selection procedures as the data.
An elongation of  X$_{\rm MAX}$ is shown in       
Fig.\ref{xmax-elongation.label}
together with expectations for protons and irons
using various hadronic interactions models:
QGSJET-II-03, QGSJET-01c and SYBILL 2.1.

The measured average value of  X$_{\rm MAX}$ at 10$^{19}$ eV 
is 751~$\pm$ 16.3(sys) $\pm$ 9.4(stat) g/cm$^2$ and 
the elongation rate is 24.3 $\pm$ 3.8(sys) $\pm$ 6.5(stat) g/cm$^2$.
Assuming a purely protonic composition, taking into account all
reconstruction and acceptance biases and using the QGSJET-II-03 
model, we would expect the average X$_{\rm MAX}$ 
at 10$^{19.0}$eV to be 763 g/cm$^2$ and the
elongation rate to be 29.7 g/cm$^2$ per energy decade.
We conclude that the measured composition above 10$^{18.2}$eV
is consistent with pure proton simulation: the same results as the stereo
analysis using FD(BRM) and FD(LR).
It is inconsistent with the pure iron simulation up to
10$^{19.2}$~eV, but above this energy the statistics
are not enough to separate proton and iron.
Considering the uncertainty of the hadronic
interaction models, our results
may not exclude the composition being
light nucleus in this energy range. 

%%%%%%%% -----------------------------------------------------
\begin{figure}[h]
\begin{center}
  \includegraphics[width=0.9\linewidth]
  {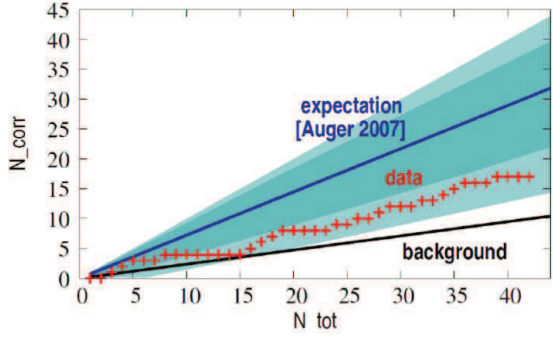}
  \caption{
Red crosses show the time development of AGN correlation
in the TA SD data.
The abscissa is a cumulated number of events with energies
57~EeV~$<$~E and $\theta$ $<$ 45.0$^\circ$,
the ordinate is the number of events correlated with the AGN
within 3.1$^\circ$. The AGNs with z $<$ 0.018 were taken 
in the VCV catalog as in the AGN correlation search
by PAO. The blue line and regions represent 
the expectation and uncertainty from the PAO measurement
\cite{auger-agn-correlation}.
}
  \label{anisotropy-agn-timedev.label}
\end{center}
\end{figure}
\begin{figure}[h]
\begin{center}
\includegraphics[width=1.0\linewidth]
  {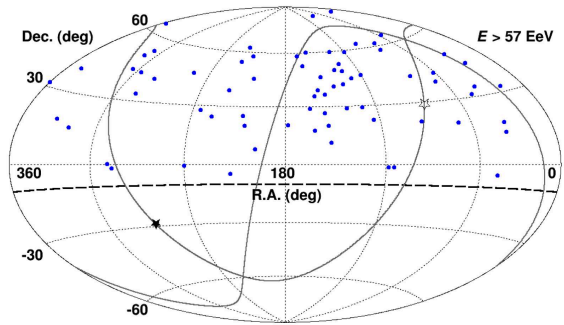}
  \caption{Sky plot of events with energies
57EeV $<$ E and $\theta$ $<$ 55$^\circ$
in equatorial coordinates. The 5-year SD data
is plotted. 
Galactic Plane (GP), Galactic Center (GC) and Super-Galactic
Plane (SGP) are indicated. The broken line shows
the limit of TA acceptance at decl. = $-$10$^\circ$. 
}
  \label{hotspot-skymap-5y.label}
\end{center}
\end{figure}

\section{Anisotropy}
The PAO had shown in 2007 that arrival directions
of UHECRs are correlated with AGNs in the VCV catalog
in the southern hemisphere
\cite{auger-agn-correlation}. 
The most significant correlation
was obtained for events 60~EeV $<$ E,
associating within 3.1$^\circ$ with
the AGN less than 75~Mpc away (or z~$<$~0.018).
After the best fit condition was determined,
an independent sample of events
confirmed the chance probability was 1.7~$\times$~10$^{-3}$.
For the combined set of events with updated calibration,
the largest correlation was obtained for events with 57~EeV $<$ E.

\begin{figure}[tbh]
\begin{center}
\includegraphics[width=1.0\linewidth]
  {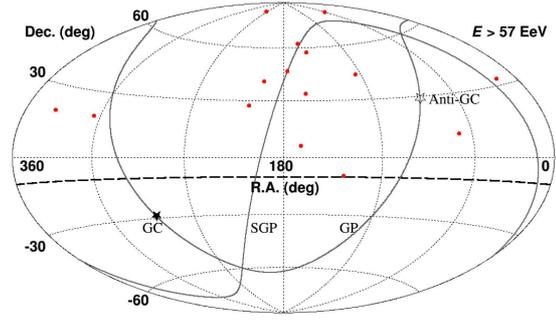}
  \caption{Skyplot of events in the 6th year only}
  \label{hotspot-skymap-6th-year.label}
\end{center}
\end{figure}
\begin{figure}[tbh]
\begin{center}
  \includegraphics[width=0.95\linewidth]
  {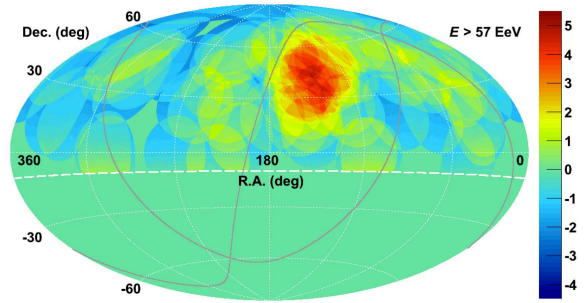}
  \caption{Significance of flux enhancement.
Oversampling with 20$^\circ$ radius circle is made for
events in 
Fig.\ref{hotspot-skymap-5y.label}.
The random background was
calculated by throwing 100k sets of 72 isotropic MC events.
The maximum significance is reached at R.A.=146.7$^\circ$
and decl.= 43.2$^\circ$. The Li-Ma significance 
of this spot is 5.1~$\sigma$.}
  \label{hotspot-significance.label}
\end{center}
\end{figure}

We repeated the correlation analysis with AGN
in the northern hemisphere
using the same selection criteria and the same catalog
as the PAO measurement.
Fig.\ref{anisotropy-agn-timedev.label}
shows the chronological progress of the number of correlated events
collected in 5 years of TA SD runs. As of May 2013,
we obtained 42 events above 57~EeV and 17 events were correlated with
the AGNs within 75~Mpc whereas we expect $\sim$10 events are
accidentally associated with 472 AGNs in the catalog
if the events are randomly distributed over the sky. The
probability of 17
or more events correlated by chance was 1.4\%
\cite{ta-agn-correlation}. 
This analysis used the same sample as the SD spectrum analysis
with tight cuts: $\theta$ $<$ 45.0$^\circ$ and 
the shower core to be more than 1200~m   
inside the array border.

In order to quantify the significance of
a possible flux enhancement seen in 
the highest energy sample of 57~EeV $<$ E
off the Super Galactic Plane 
\cite{tinyakov-rio-icrc},
we loosened the event selection to accept
all events with $\theta$ $<$ 55$^\circ$ 
and with the core location inside the array border.
It  increased the sample size
above 57~EeV from 42 events to 72 events
in the 5-year sample for the ``hotspot'' analysis
(Fig.\ref{hotspot-skymap-5y.label}).
The looser cuts deteriorated 
the angular resolution from 1.0$^\circ$
to 1.7$^\circ$; the energy resolution
became from 15\% to 20\% according to
the simulation.

We made an oversampling of the 72 events
with a circle of radius 20.0$^\circ$
in the sky plot.
An average of expected number of events 
from the isotropic arrival direction
was calculated for each point of the sky
by producing 100k sets of 72 MC
events uniformly thrown over the sky
(exposure $\propto$ sin($\theta$)~cos($\theta$)).
The maximum deviation from the isotropic expectation
was then searched over the sky with 0.1$^\circ$ step.

The obtained significance map is shown in
Fig.\ref{hotspot-significance.label}. 
The largest deviation of 5.1~$\sigma$ occurred
at R.A.=146.7$^\circ$
and decl.= 43.2$^\circ$,
where 19 data events were found 
within the circle of r~=~20$^\circ$,
and 4.49 events were 
expected from the isotropic arrival distribution
\cite{ta-hotspot}.

We estimated a chance probability of this ``hotspot'' to appear
from the isotropic arrival distribution by simulation:
1M sets of 72 events with isotropic distribution
were generated and analyzed in the same way
as the data by searching for the sky direction 
to give the maximum deviation. In the search for the maximum
deviation, we not only searched with r=20$^\circ$
but also included the searches with
r = 15$^\circ$, 25$^\circ$, 30$^\circ$, 35$^\circ$.
This is to compensate a potential ``eye-ball scan''
we might have made unintentionally
in selecting the radius of 20$^\circ$. 

Out of 1M sets simulated,
we found 37 sets whose maximum significance exceeded
5.1~$\sigma$.
The chance probability
of the hotspot appearing from the isotropic distribution
somewhere in the TA's acceptance
(direction of hotspot)
and with the radius of 15$^\circ$ - 35$^\circ$
(size of hotspot)
was determined to be 3.7$\times$10$^{-4}$ or 3.4$\sigma$.

We repeated the same analysis for the data collected
in the 6th year, May 2013 $-$ May 2014. We
obtained 15 additional events above 57~EeV
by the same loose cuts
(see Fig.\ref{hotspot-skymap-6th-year.label}) 
and 4 events are found in the hot spot determined by
the 5-year data sample. 
%The corresponding chance probability
%is 7\% from the 6th year only.
For all the data collected in 6 years, the significance
of the hotspot increased to 5.55$\sigma$
and the chance probability for getting
more than 5.55$\sigma$ from the isotropy advanced to
4.0~$\sigma$ (from the 3.4~$\sigma$ of the 5-year data).

%%%%%%%% ------------------------------------------------
\section{Prospects}
\begin{figure}[tbh]
\begin{center}
\includegraphics[width=0.8\linewidth]
{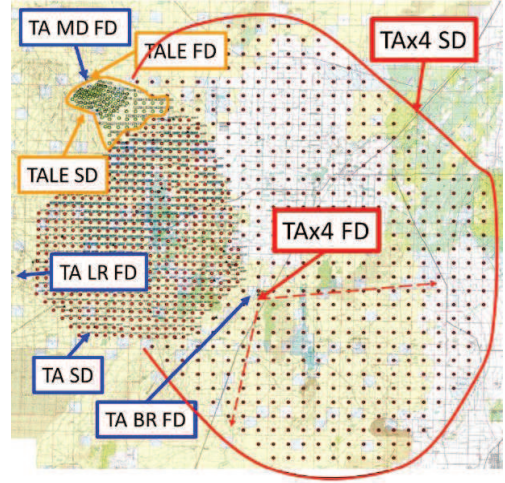}
\caption{
Planned detector layout of TA$\times$4. Additional
SDs and FDs will be installed on the east side of the existing TA.
The SD area will be quadrupled for the highest energy events.
}
\label{TAx4-map.label}
\end{center}
\end{figure}

TA plans to extend the acceptance at the highest energy region
by adding 500 additional SDs and 1 or 2 more FD stations.
The plan is called
TA$\times$4 and the detector layout 
is shown in Fig.\ref{TAx4-map.label}.
The SDs of the same design will be deployed
with a larger spacing of 2.1km extending east of
the existing TA SD array. 
The new array will be 96\% efficient at 57~EeV
for the trigger and reconstruction, and the SD acceptance
will be quadrupled to $\sim$3000~km$^2$ above this energy.
The efficiency at 10$^{19}$~eV is expected to be
26\%. The new FD stations will be
composed of refurbished HiRes telescopes with
FADC readout already used for the TALE FD telescopes. 

The construction of TA$\times$4 will take 2 years,
and the corresponding budgets have been applied
in Japan, USA, Korea and Russia. After 3 years of
operation in 2018-2020, TA$\times$4 will accumulate
21-year equivalent of TA SD data and 18-year equivalent of
TA hybrid data, if the construction is approved in 2015.
With $\sim$250 events collected above 57~EeV by 2020,
the hotspot may be established at 7~$\sigma$ or higher
and properties of the hotspot such as the energy spectrum,
the composition and finer sub-structure will be studied.
There may be other weaker hotspots appearing
in the northern sky. The X$_{\rm MAX}$ information
will be available in the cutoff region,
and the cutoff shape will be studied in detail with respect to the
expectation of the GZK cutoff
\cite{tax4-springdale}.

%%%%%%%% -----------------------------------------------------
\section{Summary}
In this conference, we presented the energy spectrum and
the anisotropy measured by TA using 6 years of SD data. 
We also presented X$_{\rm MAX}$ results analyzed 
in the hybrid mode. 

The observed spectrum above 10$^{18.2}$eV
shows a dip at 10$^{18.7}$eV and a cutoff at 10$^{19.7}$eV.
These features are well reproduced by the model calculation
of the GZK process: energy loss of extra-galactic protons
by the interaction with the CMB and IR background. The
simulation suggested a source spectrum of
E$^{-2.2}$ and source evolution of (1+z)$^{6.4}$ for
z~$<~$0.7.

A comparison of X$_{\rm MAX}$ distribution
with model simulations (QGSJET-II-03), we showed 
the primary composition is consistent with 100\% proton and
inconsistent with 100\% iron for energies
10$^{18.2}$~eV $<$ 10$^{19.2}$~eV.
Due to the ambiguities arising from hadronic interaction models, 
we can not refute an inclusion of light nuclei in the
same energy range. The energy region above 10$^{19.2}$ requires
more statistics for a better suggestion of the composition.

Above the cutoff energy, 57EeV~$<$~E, 
we observed a hotspot near the direction
of  Ursa-Major with a size of r=20$^\circ$. The probability of
this hotspot being formed by chance from the isotropic
distribution is 4~$\sigma$. The center of the hotspot
is 19$^\circ$ off from the Super-Galactic Plane,
and there is no immediate candidate of the UHECR source
nearby. 

Further studies of UHECRs at TA will continue with an extension
project TA$\times$4, which will quadruple the SD acceptance
and double the hybrid acceptance at the highest energy 
by 2017. 

%%%%%%%% -----------------------------------------------------

\section*{Acknowledgement}
%\vspace{10mm}
The Telescope Array experiment is supported 
by the Japan Society for the Promotion of Science through
Grants-in-Aid for Scientific Research on Specially Promoted Research (21000002) 
``Extreme Phenomena in the Universe Explored by Highest Energy Cosmic Rays'', 
and the Inter-University Research Program of the Institute for Cosmic Ray 
Research;
by the U.S. National Science Foundation awards PHY-0307098, 
PHY-0601915, PHY-0703893, PHY-0758342, PHY-0848320, PHY-1069280, 
and PHY-1069286 (Utah) and 
PHY-0649681 (Rutgers); 
by the National Research Foundation of Korea 
(2006-0050031, 2007-0056005, 2007-0093860, 2010-0011378, 2010-0028071, R32-10130);
by the Russian Academy of Sciences, RFBR
grants 10-02-01406a and 11-02-01528a (INR),
IISN project No. 4.4509.10 and 
Belgian Science Policy under IUAP VI/11 (ULB).
The foundations of Dr. Ezekiel R. and Edna Wattis Dumke,
Willard L. Eccles and the George S. and Dolores Dore Eccles
all helped with generous donations. 
The State of Utah supported the project through its Economic Development
Board, and the University of Utah through the 
Office of the Vice President for Research. 
The experimental site became available through the cooperation of the 
Utah School and Institutional Trust Lands Administration (SITLA), 
U.S.~Bureau of Land Management and the U.S.~Air Force. 
We also wish to thank the people and the officials of Millard County,
Utah, for their steadfast and warm support. 
We gratefully acknowledge the contributions from the 
technical staffs of our home institutions as well as 
the University of Utah Center for High Performance Computing (CHPC). 
%

%%%%%%%% -----------------------------------------------------
%\vspace{10mm}

\end{document}